\documentclass[aps,prd,preprintnumbers,
superscriptaddress,nobalancelastpage,nofootinbib,floatfix]{revtex4}

\usepackage{graphicx}
\usepackage{amssymb,amsmath}

\begin{document}
\title{Developing the Theory of Flux Limits from $\gamma$-Ray Cascades}
\author{John A. Cairns}
\email{cairnsj@mps.ohio-state.edu, john@2ad.com}
\affiliation{Department of Physics, The Ohio State University,
Columbus, OH 43210, USA}
\affiliation{Center for Cosmology and Astro-Particle Physics,
  Department of Physics, The Ohio State University, Columbus, OH
  43210, USA} 

\date{\normalsize May 18, 2007}

\begin{abstract}
  Dark matter annihilation and other processes may precipitate a flux
  of diffuse ultra-high energy $\gamma$-rays.  These $\gamma$-rays may
  be observable in present day experiments which observe diffuse
  fluxes at the GeV scale.  Yet the universe is presently opaque to
  $\gamma$-rays above 10~TeV.  It is generally assumed that cascade
  radiation is observable at all high energies, however the disparity
  in energy from production to observation has important consequences
  for theoretical flux limits.  We detail the physics of cascade radiation
  development and consider the influence of energy and redshift
  scale on arbitrary flux limits that result from electromagnetic cascade.

\end{abstract}

\maketitle\thispagestyle{empty}



\pagestyle{headings}

\section{Introduction}

{\leavevmode
\begin{figure}[h!]
\centering
\includegraphics[width=3.25in]{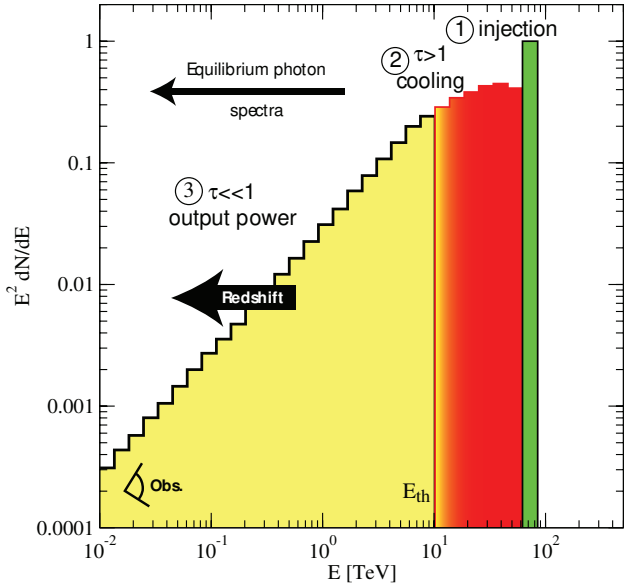}
\hskip 0.5in
\includegraphics[width=3.0in]{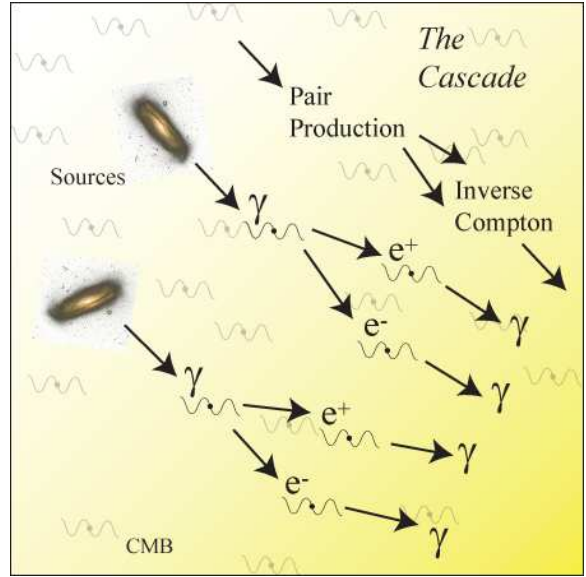}
\caption{The pair-photon cascade cycles $\gamma$-rays through a series
  of steps that cool these $\gamma$-rays while conserving energy overall.  The
  total injected power is observable in output spectra.  First,
  diffuse isotropic sources inject $\gamma$-rays above the threshold
  energy (1).   Next these $\gamma$-rays cool through pair production and
  inverse Compton scattering (2).   Finally, the cooled $\gamma$-ray
  spectrum is observable as a spectrum of inverse Compton photons (3).}
\label{Fig:TheCascade}
\end{figure}}

The GZK process produces a flux of ultra high-energy $\gamma$-rays.
These gamma rays are scattered by the CMB.  Furthermore, recent work
concerning dark matter annihilation into $\gamma$-rays suggests this
is an important pathway for constraining and revealing the nature of
dark matter particles \cite{Ullio:2002pj}.  Evidently the universe is
replete with non standard model particles.  These particles must
account for roughly 90\% of the matter budget for the universe as a
whole.  While the galaxy scale distributions of these particles remain
in contention, it is known that on large scales dark matter particles
are distributed with relative uniformity.  Any uniform distribution of
particles that annihilate to $\gamma$-rays are subject to substantial
observational constraints in both existing and planned experiments.  A
caveat occurs when these particles are produced above the threshold
energy for pair production at cosmological scales.  The flux of
particles at Earth may be mitigated by cosmogenic propagation.
Therefore, before one may constrain particle fluxes based on
electromagnetic cascades, one must have a detailed understanding of
cascade development, energy scale, production time scale and detector
sensitivity.

In addition, the origin of the diffuse extragalactic component of
EGRET observations is a deeply held mystery in $\gamma$-ray
astrophysics \cite{Bergstrom:2001jj, Taylor:2002zd, Strong:2004ry}.
In the galaxy it is predicted that inverse Compton scattering of
cosmic-ray electrons on interstellar photons may be the source of
diffuse $\gamma$-ray emission \cite{Strong:1998pw, Moskalenko:1998gw}.
Alternatively, both cosmological and exotic ultra-high energy
processes may contribute a portion of the extragalactic $\gamma$-ray
spectrum.  Fermi shock acceleration may inject ultra-high energy
cosmic-ray electrons beyond the TeV scale \cite{Gaisser:2006ne}.  All
of these extragalactic high-energy processes are rare and diffuse.  If
these processes contribute $\gamma$-rays above the pair production
threshold, then the determination of the observable spectra resulting
from diffuse isotropic cosmological injection is an essential
precursor to considering astrophysical models.

\section{The Cascade}

An isotropic injection of ultra-high energy $\gamma$-rays is subject
to the pair-photon cascade.  The cascade was suggested as early as
1948 \cite{Feenberg:1948}, however it was described in detail by
Bonometto and Rees in 1971 \cite{Bonometto:1971}.  This process
results when the coupled chain reaction of pair production
\eqref{PairProduction} and inverse Compton scattering \eqref{ICS} is
possible.

\begin{equation}
\label{PairProduction}
\gamma \gamma \rightarrow e^+e^-
\end{equation}

\begin{equation}
\label{ICS}
e\gamma \rightarrow e\gamma
\end{equation}

Berezinsky illustrates the basic argument in several works
\cite{Berezinsky:1975, Berezinsky:1990, Berezinsky:1999az}.  First, an
incoming high energy $\gamma$-ray encounters a thermal background
photon and forms an $e^+e^-$ pair.  Pair production is the subject of
a useful review by Motz, Olsen, and Koch \cite{Motz:1969}.  At low
energies, with s-wave scattering, the outgoing electron and positron
share the incoming total energy:

$$E_e = E_\gamma/2$$

At high energies, a leading particle carries away most of the incoming
energy.  This is the forward scattering limit.  The threshold for pair
production presents an absolute lower bound on the energy where the
cascade may occur.  This threshold depends on the average energy of
isotropic target photons, $\epsilon_0$, and the mass of the electron.
At present this value is:

$$E_{th} = m_e^2/{\epsilon_0} = 10~\left( {\epsilon_0\over {5\times10^{-3}~{\rm
      eV}}}\right)~\,{\rm TeV}$$

At this energy the target distribution is given by the infrared
background, while at PeV leading energies the cosmic microwave
background may participate.  For linear cascades, which develop in low
density regions of the universe, the products of cascade development
do not participate directly, therefore it is safe to assume that the
target distribution is mono-energetic, as in the following brief
discussion.  Later the thermal photon spectra is presented along with
the kinetic solution.  Stawarz and Kirk discuss non-linear cascades in
a recent work \cite{Stawarz:2007hu}.

In all of the past discussions of cascade development, one crucial
factor is typically under represented.  The background distributions
of target photons typically evolve rapidly with redshift, $\epsilon
= \epsilon_0\,{(1+z)}$ \cite{Stecker:2005qs}.  We will see that
redshift plays an important role in observability of cascade fluxes.

After pair-production occurs on target photons, the resulting stream
of high energy electrons and positrons are susceptible to inverse
Compton scattering.  Through inverse Compton scattering the energy of
these electrons or positrons is transferred into a background
photon which results in a new high energy outgoing $\gamma$-ray.  The
electrons and positrons are left behind and do not participate
further.  The vast majority of the incoming energy is carried by the
outgoing $\gamma$-ray.  In the low energy limit, the average energy
loss fraction is \cite{Gould:1975}.

\begin{equation}
\label{Eq:ComptonEFrac}
 f = {\Delta E \over E} \simeq {4\over 3} {{E\epsilon_0} \over
  m_e^2}
\end{equation}

An outgoing electron or positron resulting from $\gamma$-ray injection
at threshold will have energy:

$$ E_e = {1 \over 2 } m_e^2/\epsilon_0 $$

Thus, the Cascade will exhibit a transition energy when outgoing
$\gamma$-rays no longer have sufficient energy to initiate pair
production.  This transition energy, called the {\em critical
energy}, $E_c$, is the energy of an outgoing inverse Compton scattered
$\gamma$-ray resulting from an electron or positron produced with
energy $E_{th}/2$.

$$ E_{\gamma, out} = f E_e$$

or

$$ E_c = E_{\gamma,out} = {4 \over 3} E_e^2\epsilon_0/m_e^2 = {1\over 3}
E_{\gamma, th} $$

At present, the cascade begins to transition from recycling to emission
above 3~TeV.  This process is depicted in Fig.~\ref{Fig:TheCascade}.  The
net effect of the cascade is the reprocessing of one
incoming high energy $\gamma$-ray into pairs of outgoing
$\gamma$-rays, each with roughly half of their parent's energy.  These pairs
form others until numerous final particles result.   All of these
final particles combined share the energy of the incoming $\gamma$-ray.

There are two important energy scales for cascading particles.  Below
the energy of transition and above.  Above $E_c$, energy is conserved
by the cascade process, therefore we may immediately write the
spectrum, $E^2 dN/dE = const$.  Below $E_c$, the last generation of
$\gamma$-rays are unable to pair produce, therefore they must escape
after inverse Compton scattering which has no low energy threshold.
These $\gamma$-rays lose energy and conserve number.  A deduction of
this spectrum is given in App.~\ref{App:BerezinskySpectrum}.

\begin{equation}
\label{Eq:PreviousCascadeSpectra}
{dN\over dE} = \begin{cases}
A E^{-2} & E > E_c \\
A^\prime E^{-3/2} & E < E_c \\
\end{cases}
\end{equation}

This conventional approximation to the cascade is prevalent in many
studies of $\gamma$-ray fluxes.   The analysis presented considers the
effect of one step of the cascade and allows for reasonable
approximations to be made for given models.  It is also very simple to
consider the effects of two cascade steps.   This derivation is given
in App.~\ref{App:BerezinskyTwoStep}.   Finally, for a multi-stage
cascade the following approximation is appropriate:

\begin{equation}
\label{Eq:CascadeSpectra}
{dN\over dE} = \begin{cases}
A E^{-2} & E > E_{th} \\
A^\prime E^{-3/2} & E_c < E < E_{th} \\
A^{\prime\prime} E^{-1} & E < E_{c} \\
\end{cases}
\end{equation}

\noindent The cascade conserves total energy as it cools the
incoming $\gamma$-rays and one exploits this to fix the relation between
injection spectra and emission spectra and to determine the
normalization constants.

This discussion is essentially complete less one crucial detail.
While present observations exist at GeV energies, the present value of
$E_c$ is about 10~TeV.  In order to develop limits on particle fluxes
we must consider this issue.  The opacity of the universe evolves in
redshift.  Present experiments will have sensitivity to
electromagnetic cascades which develop at $z\approx 1$ and beyond.  If
cascade radiation is produced only at present, it may only connect
with present diffuse observations through extreme downscattering.

All cascade flux limits are deduced from the assumption that the
energy density of cascade radiation does not exceed experimental
sensitivity as in Fig.~\ref{Fig:EGRETvsCascade}.  For an energy
density $\rho$ [GeV/cm$^3$], the following relation is true
today:

\begin{equation}
\label{omegalimit}
\rho_{cas} < \rho_{obs}
\end{equation}

\noindent The flux limit on particles injected above $E_c$ directly
follows from this relation: 

$$\left(E^2 {d\Phi\over dE}\right)_{in} < \, \left(E^2 {d\Phi\over
  dE}\right)_{obs} = {c \over {4\pi}} \rho_{obs} $$

\noindent The most important relevant
observation is from EGRET, see Fig.~\ref{Fig:EGRETvsCascade}.  The
EGRET experiment on-board the Compton $\gamma$-ray observatory has set
an upper bound at 30~GeV of roughly
$10^{-6}$~GeV~cm$^{-2}$~s$^{-1}$~sr$^{-1}$ \cite{Strong:2004ry}. 

\vskip .25in
{\leavevmode
\begin{figure}[t]
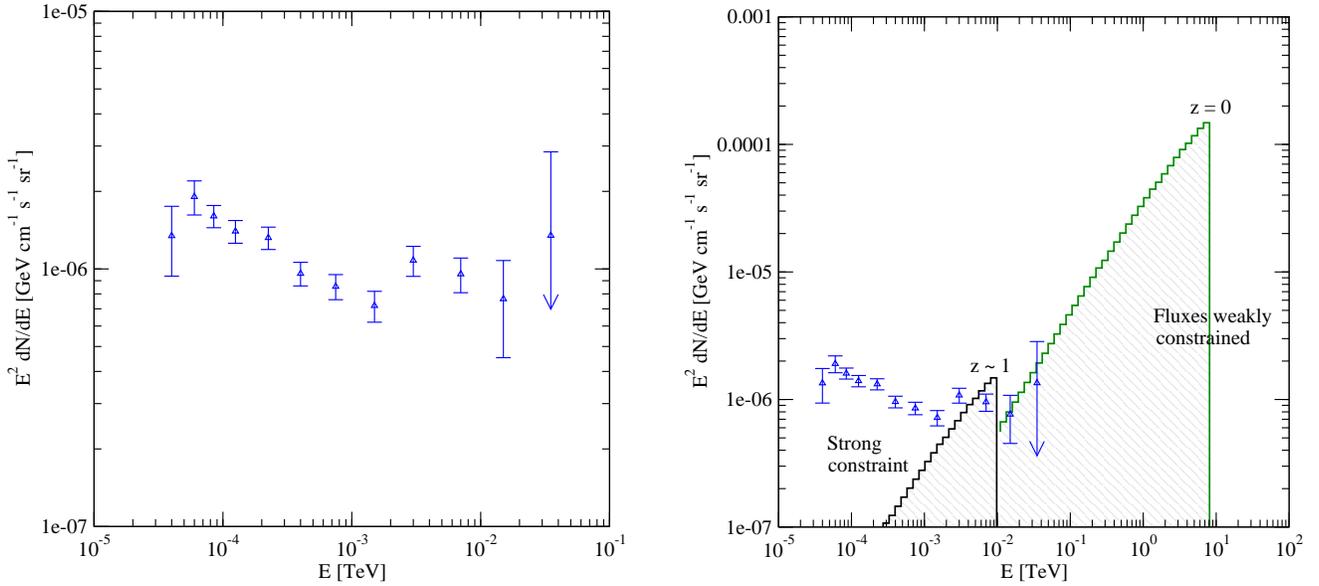

\centering
\includegraphics[width=3.25in]{EGRET.eps}
\hskip .25 in
\includegraphics[width=3.25in]{cascadevsEGRET.eps}
\caption{EGRET is sensitive to extragalactic $\gamma$-ray backgrounds
  below 30~GeV (left) \cite{Strong:2004ry}.  When compared with fluxes
  of cascading particles it is evident that experimental limits made at present
  apply only above $z\approx 1$ (right).}
\label{Fig:EGRETvsCascade}
\end{figure}}

Therefore there are two ways to develop cascade limits based on
observations at present.  Either cascade limits must be deduced from a
prior epoch when the universe was opaque to $\gamma$-rays at 30~GeV,
i.e. $z \approx 1$, or limits must incorporate the emission spectrum
of $\gamma$-rays below $E_c$ to decide what part of the present era
flux contributes to a particular experiment.

First, it is straight-forward to predict the redshift effects on cascade
radiation.  Consider, the cascade radiation in a comoving volume ${\cal
  V}$ defined for an arbitrary but fixed number of cascade particles.
Then by \eqref{omegalimit}, $\rho_{cas}{\cal V} < \rho_{obs}{\cal V}$
is also true.  However, ${\cal V}$ may be parameterized by a scale
factor $a(t)$, and coordinate representation $F({\bf r})$, ${\cal V} =
4/3\pi a^3 F^3$.  Since $a$ parameterizes all time dependence and
similarly $F$ all spatial dependence one may easily conclude that,
$\rho_{cas} a^3 < \rho_{obs} a_0^3$ is always true.  Here $a_0$ is
used to reference the present day value of the scale factor.
This is more customarily written in terms of redshift:

$$ \rho_{obs} > \rho_{cas} ({a\over a_0})^3 = {\rho_{cas} \over
  (1+z)^3} $$

\noindent These redshift effects are generally canceled by source
  evolution with density, i.e. $n = n_0 (1+z)^3$.
  
  Alternatively, cascade radiation may be contributed by
  downscattering of final stage cascade electrons and positrons.  We
  will treat this spectral development in detail below, yet one can
  make an initial estimate by taking inverse Compton scattering to
  have an $E^{-3/2}$ spectrum.  By \eqref{omegalimit}, for emission
  below $E_c$ and observation at $E_{obs}$ one has $\rho_{obs}
  N_\gamma > \rho_{cas} N_\gamma $ or $\rho_{obs} E_{obs}^{-3/2} >
  \rho_{cas} E_c^{-3/2}$.  As in the case of EGRET observations, if
  $E_c/E_{obs} \approx 100$, then the correction factor will be
  $1/1000$.  Even GLAST will only reduce this to $1/100$.
  Unfortunately, present day cascade radiation will almost certainly
  not contribute to present day experimental observation through this
  mechanism.  Eq.~\eqref{omegalimit} must be extended with the
  following relation:

\begin{equation}
\label{omegalimiteqn}
\rho_{cas}^\prime(E) = \int_{E}^{E_{max}}{dE^\prime \int_{0}^{\infty}{ dz
    {{\rho_{cas}(E_c(z) = E^\prime)} \over {(1+z)^3}} \left( {E_c \over E^\prime} \right)^{-3/2}}}
\end{equation}

From this one may directly consider a few simple cases:

\begin{enumerate}
\item Present era high energy injection and observation at $E_c$: In
  this case $\rho_{cas}^\prime = \rho_{cas}(E_c)$.  
\item Present day high energy injection and observation somewhat below
  $E_c$.  In this case one has a $(E_c/E_{obs})^{-3/2}$ penalty:
$$ \rho_{cas}^\prime = \rho_{cas} \left( {E_c\over E_{obs}} \right)^{-3/2}$$
For the currently practical case of EGRET, contributions to cascade limits
by these fluxes are ruled out by a factor of $1/1000$.  This could
only result in very weak limits on particle fluxes.
\item Observation of a 30~GeV process with significant contributions
  at $z=1$.   This might be considered the 'standard' or conventional
  case.  
  Here the best case contribution is:
$$ \rho_{cas}^\prime = {\rho_{cas, z=1}\over 2^3} $$
\end{enumerate}

Consideration of other processes requires a more detailed discussion,
however it is already evident that $\gamma$-ray cooling has important
consequences for cosmological production processes.  Any $\gamma$-rays
injected above $E_{th}$ may be observable below this energy.  The
source of the diffuse flux of $\gamma$-rays below 30~GeV is an open
question for astrophysics.  However, The energy resulting from cascade
$\gamma$-rays is narrowly distributed around $E_c$ with little or no
contribution as energies approach 30~GeV.  These fluxes are suppressed
by three orders of magnitude.  Important cosmological processes would
have to inject enormous amounts of energy above $E_{th}$ to even
partially contribute to EGRET diffuse observations today.  Therefore
it is unlikely that the source of the EGRET diffuse observation is
local $\gamma$-ray injection above $E_{th}$.

\section{Saturated Pair Cascades and Differential Fluxes}

Let us now turn to the saturated pair-photon cascade problem to
develop a more detailed spectral understanding of cascading particles.
For cosmogenic processes one is only concerned with saturated
propagation.  High energy injection processes are defined to be
saturated if through extremes of cosmological scale or density all
injected $\gamma$-rays eventually must pair produce
\cite{Svensson:1987}.  In short, the universe is completely opaque to
these particles.  Obviously, this calculation is not sensitive to
conditions in which the universe is not completely opaque.  The
cascade does not conserve particle number so repeated solution of
scattering relations is required to determine a final emission
spectrum.

To solve saturated propagation it is customary to employ the method
proposed by Guilbert \cite{Guilbert:1981}.  An integral kinetic
equation gives the spectra resulting from sources.  The cross sections
of $\gamma$-ray scattering are well known.  It is possible to directly
calculate the probability of a $\gamma$-ray scattering from a given
energy to any other energy.  Repeated convolution of the scattering
probability with the incident number distribution finally determines
the resulting spectrum.

Svensson and Zdziarski \cite{Zdziarski:1985, Svensson:1987,
  Zdziarski:1988, Zdziarski:1989, Zdziarski:1989a, Zdziarski:1989b,
  Svensson:1990, Nath:1990} considered the cascade problem extensively
using a similar analytical approach.  The steady-state solution to a
kinetic equation of propagation allows an analytical determination of
final spectra based solely on rate and injection models.  If electron
escape is neglected, then the steady state electron distribution is
completely described by electron production and energy loss.

This solution applicable to diffuse cosmological processes makes
the following assumptions.

\begin{enumerate}
\item A narrow isotropic distribution of $\gamma$-rays is injected above
  $E_{th}$.   This may similarly be solved for power laws and other
  distributions.
\item The interaction scale for photons and electrons is short
  (10~Mpc) in comparison with propagation scales (Gpc).  Therefore,
  cosmogenic injection processes are saturated above the
  pair-production threshold, $E_{th}$.

\item The universe is pervaded by an isotropic distribution of soft
  background photons at a recent epoch, $z < 0.03$.  This solution may
  be extended to higher redshifts without difficulty.
\item The energy loss time-scale for electrons is negligible in comparison
  with any possible escape time-scale.
\item Any homogeneous magnetic fields present on cosmological scales
  have a negligible effect on energy loss.
\end{enumerate}

Based on these assumptions, a self-consistent solution to the
saturated pair cascade problem in the ultra-high energy low redshift
regime follows.  The observable $\gamma$-ray spectrum below $E_c$ is
then deduced.  It turns out that the isotropic cosmological background
acts as a photon calorimeter, while the conformation of any injected
spectra are lost, total energy of injection is preserved and observable
in experiments with sensitivity near $E_c$.

\section{Interactions of Cosmological $\gamma$-Rays and Electrons}

Cosmological $\gamma$-rays and electrons may be susceptible to energy loss
through inverse Compton scattering, pair production, photon-photon
scattering, Compton scattering, synchrotron radiation and redshift.
One may briefly summarize the effects of these processes to deduce the
important propagation processes.  In the conventional notation for
dimensionless energy, one refers to photon energy with $\omega =
E_\gamma/m_e$ and electron Lorentz factor $\gamma = E_e/m_e$. 

\begin{center}
\begin{table}[h!]
\begin{tabular}{ccccc}
        \hline
{\bf Incident Particle}&{\bf Process}&{\bf Target Density [cm$^{-3}$]}&{\bf Cross
  Section [$\sigma_T$]}&{\bf $\lambda$ [Mpc]}\\
        \hline
        \hline
Gamma Rays \\
\hline
& $\gamma\gamma\rightarrow e^+e^-$ on CIB (TeV)& 0.5 & $3 / 16$ &
$10$\\
& $\gamma\gamma\rightarrow e^+e^-$ on CMB (PeV)& $410.5$ & $3 /
16$ & $10^{-2}$\\
& $\gamma e \rightarrow \gamma e$ & $10^{-7}$ & $10^{-3} $ & $10^{10}$\\\hline
Electrons and Positrons \\
\hline
& $e\gamma \rightarrow e\gamma $ on CIB (TeV)& 0.5 & 1 &
$1$\\
& $e\gamma \rightarrow e\gamma $ on CMB (PeV)& 410.5 & 1 &
$10^{-2}$\\
\hline
\label{interaction_scales}
\end{tabular}
\caption{Inverse Compton scattering and pair production are important
  cosmological loss processes for electrons and $\gamma$-rays.}
\end{table}
\end{center}

\subsubsection{Pair Production from $\gamma$-Rays}

To estimate the scattering length for $\gamma$-rays I assume
cosmological $\gamma$-rays will pair-produce on an isotropic infrared
background if $\epsilon_\gamma \approx 10$~TeV.  The center of mass
energy squared for TeV gamma-rays against infrared background photons
is $s \approx 4 \epsilon_{\gamma 1} \epsilon_{\gamma 2} \approx
10^{-13}$~TeV$^2$, the dimensionless center of mass energy is
$\sqrt{\omega_1\omega_2} \approx 1$.  The cross section is found at
the peak of the pair production rate $\sigma \approx {3 \over 16}
\sigma_T$.  The Thomson cross section, $\sigma_T$, is related to the
classical electron radius by $8\pi r_e^2/3$.  The mean interaction
length for cosmogenic pair production is about 10~Mpc.  This mean free
path is minimized for injected gamma-rays at PeV energies.  In this
case, the interaction length drops to about 10 kpc because of
the increased density of microwave background targets.

$$\lambda_{\gamma\gamma\rightarrow e^+e^-} = {1 \over n\sigma} \approx
10\,{\rm Mpc}$$

\noindent In comparison with the Hubble scale, pair production is a
primary energy loss mechanism for cosmological $\gamma$-rays with
energy above $E_{th}$. 

\subsubsection{Compton Scattering of Gamma-Rays}

Compton scattering of a cosmological $\gamma$-ray on a primordial
electron is primarily important at epochs where primordial electron
density is significant.  A TeV $\gamma$-ray and background electron
system has center of mass energy squared, $s \approx 4 \epsilon_\gamma
m_e \approx 10^{-6}$~TeV$^{2}$.  In dimensionless form,
$\sqrt{\omega\gamma} \approx 3000$, the cross section for TeV
$\gamma$-rays is $\sigma \approx 10^{-3} \sigma_T$\,--\,$10^{-5} \sigma_T$.  The density of
primordial electrons is roughly equivalent to the density of baryons,
$n_e \approx n_b \approx 10^{-7}$ cm$^{-3}$.  The mean interaction
length of Compton scattering is only significant at high redshifts.

$$\lambda_{\gamma e\rightarrow\gamma e} = {1 \over n\sigma} \approx
10^{10}\,{\rm Mpc}$$

\noindent The mean free path for Compton scattering is so large that
I take these losses to be negligible.

\subsubsection{Photon-Photon Scattering of Gamma-Rays}

Photon-photon scattering, $\gamma\gamma\rightarrow\gamma\gamma$, is an
important cosmological consideration at high redshift and low energy
scales when the universe was radiation dominated.  Svensson and
Zdziarski treat this process in detail in \cite{Svensson:1990}.  I
neglect this scattering process here under the assumption
of ultra-high energy propagation in a matter dominated epoch.

\subsubsection{Inverse Compton Scattering of Electrons and Positrons}

Cosmological electrons of TeV energies may scatter on isotropic
primordial photons.  The center of mass energy squared is $s \approx 4
E_e \epsilon_\gamma $.  The electron energy, $E_e$, is about 1~TeV.
The energy of infrared photons averages about $10^{-2}$~eV, so the
dimensionless center of mass energy is $\sqrt{\omega\gamma} \approx
0.5$.  This is well within the Thomson regime so the cross section
is the Thomson cross section.  The density of soft infrared target
photons is $0.5$ cm${^{-3}}$.  The interaction length of TeV electrons
is about one Mpc.  This is a rough estimate made for illustrative
purposes, the integration below uses exact distributions.  This
interaction length is minimized when the CMB can participate.  At
PeV energies the interaction length of an electron drops
to about 1~kpc.

$$\lambda_{e\gamma\rightarrow e\gamma} = {1 \over n\sigma_T} \approx 1
\,{\rm Mpc}$$

\noindent As a result inverse Compton scattering is a primary energy loss
mechanism for cosmological electrons above one TeV.

\subsubsection{Synchrotron Radiation from Electrons and Positrons}

Cosmological electrons in a magnetic field, ${\bf B}$, emit $\gamma$-rays
as they accelerate through a helical trajectory.  The energy loss
through this mechanism is proportional to magnetic energy density or
${\bf B}^2$.

$$ {dE \over dt} \approx {4\over 3} \sigma_T c \beta^2\gamma^2 {{\bf B}^2
  \over 8\pi}$$

\noindent The cosmological magnetic field density has been constrained
at less than $10^{-12}$~Gauss.  Since synchrotron loss depends on
${\bf B}^2$, I assume that cosmological magnetic fields may be
neglected in this calculation.  See \cite{Blumenthal:1970} for a
detailed treatment of this loss process.

\subsubsection{Redshift}

The energy loss scale due to redshift is of roughly the same order as
the Hubble length scale.

$$ \lambda_z \approx {c\over H_0} \approx 4\, {\rm Gpc} $$

\noindent The extension of this solution
to include detailed accounting of redshift is straightforward but not
required since the principal loss processes are far faster than this mechanism.

\section{Production in Astrophysical Processes}

For particles susceptible to these interactions a particular
isotropic injection has flux defined by

\begin{equation}
\label{flux}
\Phi = {c\over 4\pi}\int_{E_{min}}^{E_{max}}{dE\, {dN\over dE}}
\end{equation}

The differential number density is taken as $dN/dE$ [GeV$^{-1}$
cm$^{-3}$]. Where $dN/dE\ dE$ represents the number of particles in an
interval $(E, E + dE)$ per volume.  If $N_0$ represents the density
today, the density evolves with redshift, $z$.

$$ N = N_0 (1+z)^3 $$

The flux of particles resulting from a particular astrophysical
process is determined by the production rate.  Rate is given by
\eqref{srate}, expressed in terms of the incident particle distribution
$n$, and reaction cross section, $\sigma$.

%

\begin{align}
\label{srate}
\Gamma(E^\prime, E) = 
c\int_{\epsilon_{min}}^{\epsilon_{max}}{d\epsilon\
  n(\epsilon)\int_{E_{min}}^{E_{max}}{dE^\prime\ {d\sigma\over dE}(E^\prime,
      E)}}
\end{align}

The term ${d\sigma\over dE}(E^\prime, E)$ represents the differential
cross section for scattering from a particle of energy $E^\prime$ to
energy $E$.  Also this integral can be normalized over target particle
distributions and arbitrary directions.  This paper adopts
the notation $\mu=\cos{\theta}$ for the scattering angle and uses
$\beta$ to refer to velocity.  The following equation represents the
production from scattering of an isotropic incident distribution on a
target distribution across a spectrum of energies.

\begin{equation}
\label{avgrate}
\dot{N} = 
c\int_{\epsilon_{min}}^{\epsilon_{max}}{d\epsilon\
  {n_T(\epsilon)}\int_{-1}^{1}{d\mu\ {{1-\beta\mu}\over
    2}\int_{E_{Imin}}^{E_{Imax}}{dE_I\
    n_I(E_I)\int_{E_{min}}^{E_{max}}{dE^\prime\ {d\sigma\over dE}(E^\prime, E)}}}}
\end{equation}

In this equation $n_T$ represents the distribution of targets and $n_I$
represents the distribution of incident particles.  I use $\dot{N}$ to
represent the first derivative in time, $dN/dt$, hereafter.

\section{The Kinetic Equation}

The kinetic equation gives the observable spectrum, $N(E)$, of
particles after scattering for a given injection spectrum $N_0(E)$.
The transport of high energy $\gamma$-rays through the cosmological
medium involves catastrophic energy loss through pair production.  The
resulting stream of electrons boosts photons from the primordial
background through inverse Compton scattering.

The contributions to $\gamma$-ray energy loss are either continuous or
discrete.  The continuous radiative transfer of a propagating particle is
described as a differential equation over spatial propagation
\cite{Rybicki:1979}. 

\begin{equation}
\label{radtransfer}
{dN\over dx}(E) = -\alpha N(E) + j
\end{equation}

Where $\alpha$ is the coefficient of assumed continuous energy
loss and $j$ is the particle injection term.  Solutions to
\eqref{radtransfer} reveal the effects of continuous energy loss on an
initial spectrum, but do not describe catastrophic loss
processes.   To consider both types of loss processes one employs a
steady-state differential equation. 

The observable spectrum of $\gamma$-rays can be deduced by considering the
repeated effects of energy loss in multiple scatterings.   Since
escape is neglected, differential changes in particle flux must be stable.

The electron steady-state is described by loss due to inverse Compton
scattering, production of $e^+e^-$ pairs, and injection \cite{Svensson:1987}.

\begin{equation}
\label{ElectronState}
\dot{N}_{e} = -\dot{N}_{e,C}(E) + \dot{N}_{e,P}(E)
 + \dot{N}_{e,in}(E)
\end{equation}

Likewise $\gamma$-rays appear after inverse Compton scattering or
injection, and disappear in pair production.  I will use $n$ (lower
case) to label a spectrum of photons and $N$ to label a spectrum of
electrons.

\begin{equation}
\label{PhotonState}
\dot{n}_{\gamma} = \dot{n}_{\gamma,C}(E) - \dot{n}_{\gamma,P}(E)
 - c{dn_\gamma\over dx}(E) + \dot{n}_{\gamma, in}(E)
\end{equation}

In \eqref{ElectronState} and \eqref{PhotonState} the subscripts ``C''
and ``P'' are used to indicate the time derivative due to
inverse Compton scattering \eqref{ICS} and pair production
\eqref{PairProduction} respectively.  The radiative transfer is given
by $dn/dx$ and \eqref{radtransfer}.  The radiative transfer term may
be used to model losses from $\gamma \gamma$ absorption in dense
environments.  The terms with ``in'' subscripts represent continuous
isotropic particle injection.  Both equations are independent of
charge, therefore electron and positron losses are identical.  In
order to consider the effects of both electron and positron production
one simply doubles the electron production rate.  For the remainder of
the article I will refer to electrons only, of course in reality half
of the denumerable electron population would be physical positrons.
This is irrelevant since I only consider the resulting flux of
$\gamma$-rays.

For many physical models involving the injection of $\gamma$-rays
there are no electron sources, in this numerical integration
$\dot{N}_{e,in} = 0$, this choice is arbitrary and made for numerical
convenience, it would be equivalent to inject first generation
electrons or positrons rather than $\gamma$-rays.

Since \eqref{ElectronState} and \eqref{PhotonState} are coupled by
pair production and inverse Compton scattering, one may adopt the
point of view of either electrons or $\gamma$-rays, I choose electrons
in consideration of previous work in this field.

Now by considering the known cross sections for inverse Compton
scattering and pair production one may write a differential equation
describing electron production in terms of energy.  Inverse Compton
scattering on an isotropic distribution of background photons is given
by \eqref{ICSRate}.

\begin{equation}
\label{ICSRate}
\dot{N}_{e,C}(E) = c
\int_{\epsilon_{min}}^{\epsilon_{max}}{d\epsilon\
  {n_T(\epsilon)}\int_{-1}^{1}{d\mu\ {{1-\beta\mu}\over
    2} \int_{E_{min}}^{E_{max}}{dE^\prime\ N_e(E^\prime) {d\sigma_C\over dE}(E^\prime, E)}}}
\end{equation}

If the background photons are thermal, the density is integrated from
the Planck distribution.  The present day temperature is
$T_0=T_{CMB}=2.725$~K and $k_b$ is the Boltzmann constant.  In
general, $T$ is proportional to $T_0 (1+z)$.  The energy evolves with
redshift according to the relation $ \epsilon = {\epsilon_0 (1+z)} $.

$$ n_T(\epsilon)\, d\epsilon = {1\over\pi^2({\hbar c})^3}
{\epsilon^2\over{\exp{({\epsilon\over k_b T})}-1}} d\epsilon $$

The primarily relevant background for $\gamma$-rays at 10~TeV is the
infrared background (CIB).  I assume this background is adequately
described by a power-law and a black body at $T_{CIB} = 2725 K$.
These assumptions are consistent with recently published detailed
models of photon backgrounds \cite{Dwek:2004pp, Stecker:2005qs}.  I
arbitrarily normalize to achieve agreement with accepted observations,
see citations above for detailed discussions of these backgrounds.

The Compton energy loss rate of electrons is equal in magnitude to the
energy production of Compton $\gamma$-rays.  For a photon of energy $\omega
m_e c^2$, and an electron of energy $\gamma m_e c^2$, the exact angle
averaged scattering rate for electron disappearance is given by the
Klein-Nishina form \cite{Coppi:1990}.

\begin{align}
\label{ICSCBRate}
&\dot{N}_{e,C} (\omega, \gamma) = 
c\int_{0}^{\infty}{d\epsilon\
  n_T(\epsilon)\int_{2\gamma(1-\beta)\omega}^{2\gamma(1+\beta)\omega}{d\kappa\ N_e(\kappa){d\sigma_{KN}\over d\kappa}}}\\
&{d\sigma_{KN}\over d\kappa}(\kappa) = {{3 \sigma_T}\over
  32\gamma^2\beta\omega^2}\left[\left(1-{4\over \kappa}-{8\over \kappa^2}\right)\ln{(1+\kappa)}+{1\over 2}+{8\over \kappa}-{1\over{2(1+\kappa)^2}}\right]\notag
\end{align}

It is usual to employ a change of variables,
$\kappa=2\gamma(1-\beta\mu)\omega$, in this equation.  In the low energy
(``Thomson'') limit this rate approaches $c\sigma_T$, the speed of
light multiplied by the Thomson cross section.  The Compton rate is
depicted in Fig.~\ref{Fig:CBComptonRate}.

\vskip .25in
{\leavevmode
\begin{figure}[t]
\centering
\includegraphics[width=3.25in]{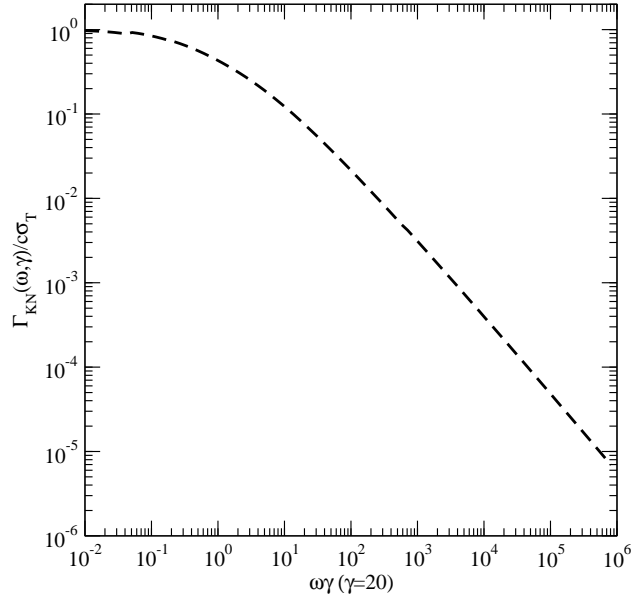}
\caption{The scattering rate of $\gamma$-rays due to the inverse Compton
  process \eqref{ICS}.   Electron energy is fixed at $20 m_e c^2$.
  This figure corresponds to Coppi and Blandford Fig.~1
  \cite{Coppi:1990}:  given here for comparison.  The Compton rate is
  unusual in that it decreases as energy increases.  The maximum rate
  in the ``Thomson'' regime occurs at $c\sigma_T$.  The rate is divided
  by $c\sigma_T$ to reflect the absolute shape of the interaction in
  dimensionless form.} 
\label{Fig:CBComptonRate}
\end{figure}}

Equation \eqref{ICSCBRate} can be analytically integrated and this is
a common step for several authors including Zdziarski, however the
resulting rate is a complicated function involving dilogarithms.
There may be no computational or intellectual benefit from performing
this integration, I omit it.  In either case one is required to
numerically integrate over the rates to deduce a final spectrum.  This
approach is also suggested by Coppi and Blandford \cite{Coppi:1990}.

To clarify the meaning of \eqref{ICSCBRate} I introduce a shorthand,
$C(E)= - \int_{E_{min}}^E{dE^\prime d\sigma/dE}$, for the portion of
the integral over \eqref{ICSRate} from $E_{min}$ up to the energy of
consideration $E$.  It is evident from Fig.~\ref{Fig:CBComptonRate}
that the derivative of the cross section is negative.

\begin{align}
\label{ICSFinal}
\dot{N}_{e,C}(E) = 
c&\int_{\epsilon_{min}}^{\epsilon_{max}}{d\epsilon\
  {n_T(\epsilon)}\int_{-1}^{1}{d\mu\ {{1-\beta\mu}\over
    2}\ }}N_{e}(E)\notag\\
&\times\left(C(E) - \int_{E}^{E_{max}}{dE^\prime\
      {d\sigma_C\over dE}(E^\prime, E)}\right)
\end{align}

In \eqref{ICSFinal} the first term represents electron energy loss due
to boosted primordial photons, it is positive (loss) because these
inverse Compton electrons are scattered to lower energies.  The second
term represents the appearance of inverse Compton scattered electrons from
higher energies.  The sign is consistent with \eqref{ElectronState}
which defines this equation as a loss rate.

The pair production or electron appearance rate is given by 

\begin{equation}
\label{PairRate}
\dot{N}_{e,P}(E) = c
\int_{\epsilon_{min}}^{\epsilon_{max}}{d\epsilon\
  {n_T(\epsilon)}\int_{-1}^{1}{d\mu\ {{1-\mu}\over
    2}\ \int_{E_{min}}^{E_{max}}{dE_\gamma\ n_{\gamma}(E_\gamma)\
    {d\sigma_P\over dE}(E_\gamma, E)}}}.
\end{equation}

The term ${d\sigma_P\over dE}(E_\gamma,E)$ represents the differential
cross section of a photon of energy $E_\gamma$ to produce an electron
of energy $E$.  Eq.~\eqref{PhotonState} reveals that there are two sources
of $\gamma$-rays that may take part in pair production, either freshly
injected $\gamma$-rays from isotropic sources or $\gamma$-rays boosted
through inverse Compton scattering.

\begin{equation*}
n_\gamma(E) = n_{\gamma, in}(E) + n_{\gamma, C}(E)
\end{equation*}

These two relations are combined in \eqref{PairFinal}.

\begin{align}
\label{PairFinal}
\dot{N}_{e}(E) &=
 \int_{\epsilon_{min}}^{\epsilon_{max}}{d\epsilon\
  {n_T(\epsilon)}\int_{-1}^{1}{d\mu\ {{1-\mu}\over
    2}\ \int_{E_{P,min}}^{E_{P,max}}{dE_\gamma\ 
 {d\sigma_P\over dE}(E_\gamma, E)}}}\notag\\
\times&\left(\dot{n}_{\gamma,
        in}(E_\gamma) +
      c\int_{E_{C,min}}^{E_{C,max}}{dE^\prime\
        N_{e}(E^\prime){d\sigma_C\over
          dE}(E, E^\prime-E_\gamma)}\right)
\end{align}

The exact photon-photon pair production rate is given by \cite{Coppi:1990}

\begin{align}
\label{CBPairRate}
&\dot{N}_{e,P} = c\int_{0}^{\infty}{d\epsilon\
  n_T(\epsilon)\int_{-1}^{\mu_{max}}{d\mu\ {{1-\mu}\over
    2}\ \int_{E_{min}}^{E_{max}}{dE_\gamma\ n_{\gamma}(E_\gamma)
    {d\sigma_{\gamma\gamma}\over dE}}}}\\
&{d\sigma_{\gamma\gamma}\over dE} = {3c\sigma_T(1-\beta^{\prime 2})\over
  16}
  \left[(3-\beta^{\prime
  4})\ln{\left({{1+\beta^\prime}\over{1-\beta^\prime}}\right)}-2\beta^\prime(2-\beta^{\prime
  2})\right].
\notag
\end{align}

Following standard notation, dimensionless parameters for photon
energy are $\omega_1=\epsilon/{m_ec^2}$ and $\omega_2=E_\gamma/{m_ec^2}$, so
$\mu_{max} = {\rm max}(-1, 1-2/{\omega_1\omega_2})$.  The electron
velocity is $\beta^\prime=[1-2/{\omega_1\omega_2}(1-\mu)]^{1/2}$.  The
pair production rate is depicted in Fig.~\ref{Fig:CBPairRate}. 

{\leavevmode
\begin{figure}[t]
\centering
\includegraphics[width=3.25in]{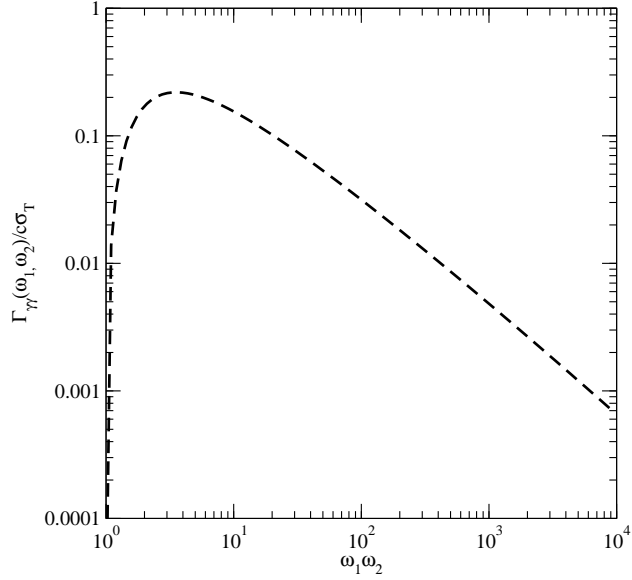}
\caption{The scattering rate of a photon of energy $\omega_1 m_e c^2$
  against a photon of energy $\omega_2 m_e c^2$ to form
  electron-positron pairs \eqref{PairProduction}.  This figure
  corresponds to Coppi and Blandford Fig.~5 \cite{Coppi:1990}: given
  here for comparison. The pair production rate peaks at about
  $3\sigma_T/16$ and then falls off as energies increase.  The rate is divided
  by $c\sigma_T$ to reflect the absolute shape of the interaction in
  dimensionless form. }
\label{Fig:CBPairRate}
\end{figure}}

For extragalactic cosmic media it is assumed that the inverse Compton
scattering and pair production mechanisms entirely describe the
cascade energy loss and no additional energy losses are present,
$dN/dx = 0$.  Combining \eqref{ElectronState}, \eqref{ICSFinal} and
\eqref{PairFinal} reveals a steady-state integral-differential
equation of electron transport.

\begin{align}
\label{ElectronTransport}
\dot{N}_{e} =
&-c\int_{\epsilon_{min}}^{\epsilon_{max}}{d\epsilon\
  {n_T(\epsilon)}\int_{-1}^{1}{d\mu\ {{1-\beta\mu}\over
    2}\ N_{e}(E)\left(C(E) - \int_{E}^{E_{max}}{dE^\prime\
      {d\sigma_C\over dE}(E^\prime, E)}\right)}}\notag\\
&+\int_{\epsilon_{min}}^{\epsilon_{max}}{d\epsilon\
  {n_T(\epsilon)}\int_{-1}^{1}{d\mu\ {{1-\mu}\over
    2}\ \int_{E_{P,min}}^{E_{P,max}}{dE_\gamma\ 
 {d\sigma_P\over dE}(E_\gamma, E)}}}\notag\\
&\times\left(\dot{n}_{\gamma,
        in}(E) +
      c\int_{E_{C,min}}^{E_{C,max}}{dE^\prime\
        N_{e}(E^\prime){d\sigma_C\over
          dE}(E^\prime, E^\prime-E_\gamma)}\right)
\end{align}

Eq.~\eqref{ElectronTransport} is comparable to Zdziarski Equation~1
\cite{Zdziarski:1988}, with solutions that show the time evolution of
the electron spectrum versus energy.  It is possible to eliminate time
dependence by considering the continuous energy loss $dE_C/dt$
\cite{Svensson:1987}.

$$\dot{N}(E) = {d\over dE}[{dE_C\over dt}N(E)] $$

Here, Zdziarski gives the continuous energy loss in terms of a small parameter
$\delta$, I take $\delta$ to be $10^{-4}$ \cite{Zdziarski:1988}:

$${dE_C\over dt}(E) = \int_{E/(1+\delta)}^{E}{dE^\prime\ (E^\prime -
  E)\dot{N}_{e,C}}$$

This alteration gives an equation comparable to Zdziarski Eq.~(2)
\cite{Zdziarski:1988}:
\begin{align}
\label{ElectronTransportFinal}
{d\over dE}[{dE_C\over dt}N_{e}] =
&-c\int_{\epsilon_{min}}^{\epsilon_{max}}{d\epsilon\
  {n_T(\epsilon)}\int_{-1}^{1}{d\mu\ {{1-\beta\mu}\over
    2}\ N_{e}(E)\left(C(E) - \int_{E}^{E_{max}}{dE^\prime\
      {d\sigma_C\over dE}(E^\prime, E)}\right)}}\notag\\
&+\int_{\epsilon_{min}}^{\epsilon_{max}}{d\epsilon\
  {n_T(\epsilon)}\int_{-1}^{1}{d\mu\ {{1-\mu}\over
    2}\ \int_{E_{P,min}}^{E_{P,max}}{dE_\gamma\ 
 {d\sigma_P\over dE}(E_\gamma, E)}}}\notag\\
&\times\left(\dot{n}_{\gamma,
        in}(E) +
      c\int_{E_{C,min}}^{E_{C,max}}{dE^\prime\
        N_{e}(E^\prime){d\sigma_C\over
          dE}(E^\prime, E^\prime-E_\gamma)}\right)
\end{align}

This equation for electron transport is a restatement of the condition
of steady state equilibrium.

$$ {d\over dE}[ N(E) {dE_C\over dt} ] = N(E) $$

In \eqref{ElectronTransportFinal} $N_e$ is present on both sides.
Also the dependence on $N_e$ is incorporated in several integral
terms.  The most effective strategy for solving
\eqref{ElectronTransportFinal} is to apply the Runge-Kutta technique.

\section{Numerical Techniques}

The Runge-Kutta technique is a method for iteratively solving a
differential equation which has the following form:

\begin{align*}
y_{n+1} = y_n + h f(x_n, y_n)\\
f(x_n, y_n) = {dy_n \over dx_n}
\end{align*}

The solution is advanced through small steps $h$, with $x_{n+1} = x_n
+ h$, and the result is accumulated.  An adaptive step size is
incorporated to reduce computation time while holding relative errors
fixed.  The C++ code utilizes 4$^{th}$ order Runge-Kutta
method with 5$^{th}$ order error checking.  It is prudent to recall
that higher-order is not synonymous with either smaller error or
improved numerical stability.  However this technique converges
rapidly for \eqref{ElectronTransportFinal}.  The details of the
Runge-Kutta calculation are described in many texts including \em
Numerical Recipes \cite{NR:2002}.\rm

I hold absolute errors to $1.0\times 10^{-4}$ at machine precision
and relative errors to $0.0$.  The equation for electron transport
gives solutions for the expected number of particles in a logarithmic
energy bin.  The initial electron spectrum is null and the initial
$\gamma$-ray spectrum is set to unit height in a logarithmic energy bin
corresponding to the injection energy.  The energy range is divided
into an arbitrary number of intervals of constant logarithmic width and
\eqref{ElectronTransportFinal} is repeatedly solved starting at higher
energies and moving to lower.

The highest energy bin must be solved first since pair production
occurs in the highest bins first, and then subsequent electrons
downscatter to lower energies over repeated iteration of the cascade
process.  In other words the electrons first appear in the higher energy bins
and move down in energy.

The result of this iteration is an electron spectrum corresponding to
the specified photon injection.  One final generation of inverse
Compton scattering reveals the outgoing $\gamma$-ray production. The
output production is integrated from the inverse Compton scattering rate to
give the final result \cite{Zdziarski:1988}.

\begin{equation}
\label{PhotonProduction}
\dot{n}_{\gamma}(E_\gamma) = \int_{E_\gamma}^{E_{max}}{dE^\prime\,
  N_{e}(E^\prime)\Gamma_{KN}(E^\prime, E^\prime-E_\gamma)}
\end{equation}

The $\gamma$-ray production is then integrated over the time variable $t$
until the output is fully saturated, i.e., until the observable energy
is equal to the injection energy.  The solution of
\eqref{ElectronTransportFinal} is depicted in
Fig.~\ref{Fig:PhotonCons}.  Finally, flux may be determined from
\eqref{flux}.

{\leavevmode
\begin{figure}[t]
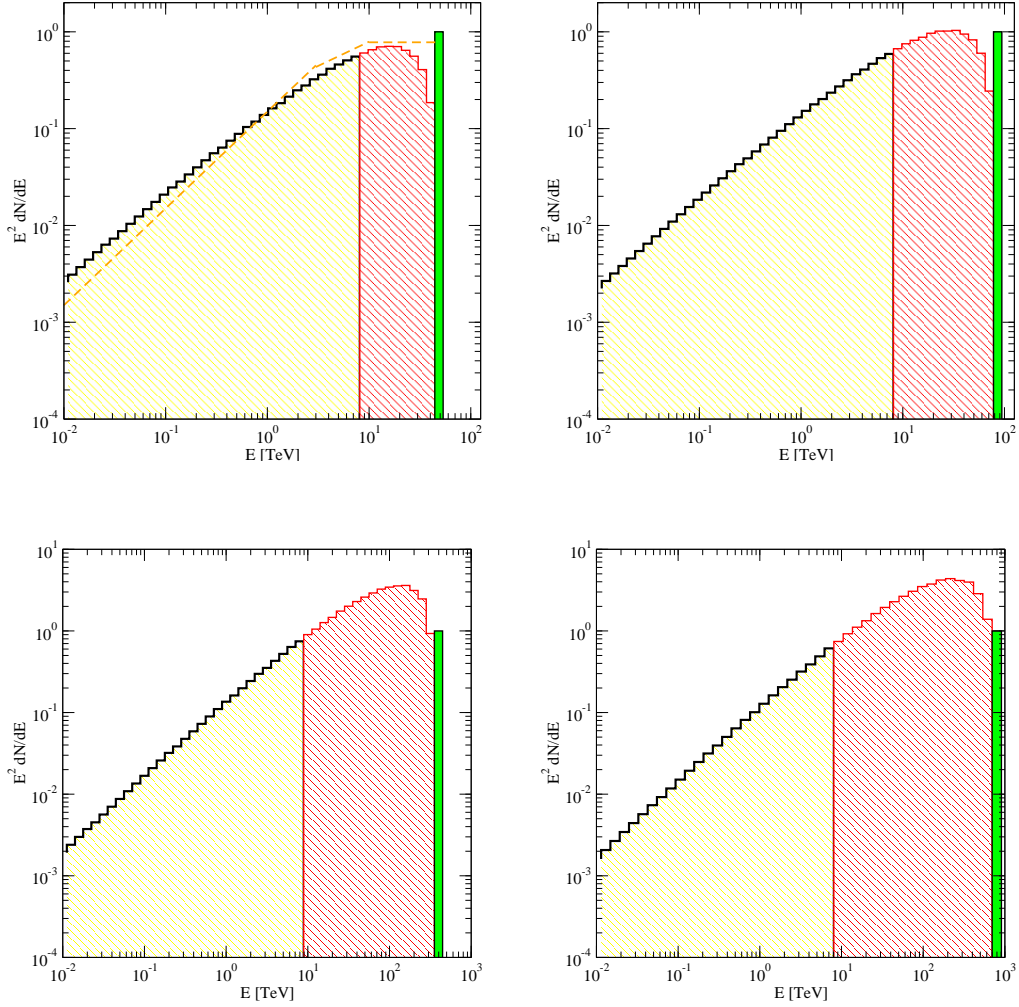

\vskip 1in
\centering
\includegraphics[width=2.5in]{cascade50TeV.eps}
\hskip 0.25in
\includegraphics[width=2.5in]{cascade100TeV.eps}
\vskip 0.375in
\includegraphics[width=2.5in]{cascade500TeV.eps}
\hskip 0.25in
\includegraphics[width=2.5in]{cascade1000TeV.eps}
\caption{A plot of \eqref{PhotonProduction} with solutions of
  \eqref{ElectronTransportFinal} as input.  The 50~TeV (top left),
  100~TeV (top right), 500~TeV (bottom left) and 1000~TeV (bottom
  right) $\gamma$-rays are injected (shown in green) above $E_{th}$.  The
  spectrum resulting from the cascade has two components.  The
  observable portion below $E_{th}$ in black and the continuously
  propagating portion above $E_{th}$ is depicted in red.  Just below the
  threshold energy of 10~TeV, the height of the resulting spectrum is
  about $1/2$ the height of the injection.
  Eq.~\eqref{Eq:CascadeSpectra} is superimposed on the 50~TeV
  injection for comparison.}
\label{Fig:PhotonCons}
\end{figure}
}

Applying numerical solution to \eqref{ElectronTransportFinal} shows
that during the cascade height (or equivalently energy) is preserved
per logarithmic energy interval on an $E^2 dN/dE$ plot as
$\gamma$-rays cool.  Fig.~\ref{Fig:PhotonCons} shows that the cascade
process recycles most of the injected energy until it finally turns
off near $E_c$, then the resulting spectrum is simply the portion due
to inverse Compton scattering.  Fig.~\ref{Fig:PhotonCons} uses a
logarithmic scale where each range represents a fixed interval of
constant energy.  Energy can be read as the height of the figure.
There are a total of 50 constant logarithmic intervals, however this
number is chosen arbitrarily to minimize processing time while
accurately representing the shape of the output spectrum.

%
%








\section{Results for Cascading Particles}

As high energy $\gamma$-rays are injected above $E_{th}$,
they immediately form pairs and begin cycling through the cascade
process, finally just above $E_c$ and after many generations,
cascading electrons and photons have nearly the same energy as the
first generation did.   Suddenly, the cascade turns off and most of
the energy is preserved in the range $E_c$ to $E_{th}$.   

Therefore, the cascade acts as a photon calorimeter.  About 85\% of
the energy injected above $E_{th}$ is ultimately observable in the energy
decade ($E_{th}/10$, $E_{th}$) near $E_c$.  However, the cascade does
not preserve the conformation of the injected spectrum.  While a
narrow bin may be injected, the observable spectrum is fixed by the
shape of the inverse Compton scattering rate and is not accurately
reflected as a simple energy translation of the input bin.

At the factor of two level, one may consider a rule of thumb.

\begin{equation}
E^2 {dn\over dE}(E_c) \approx E^2 {dn\over dE}(E_{in})
\end{equation}

This is a naive heuristic which hides the relevant physics, however
the conclusion that the majority of the injected power scatters into
the energy range at $E_c$ is valid.  Essentially this relation
demonstrates that \em the cascade\rm\ conserves energy.

Importantly the energy due to ultra-high energy $\gamma$-ray injection
processes is not lost in the sinuous cascade.  The total output power
is equal to the total input power (conservation of energy) and the
rapid ascent of the inverse Compton scattered photon spectrum ensures that
most of the observable energy will appear at $E_c$.  These
results clearly rule out a ``bin-shifting'' approach to
$\gamma$-ray energy conservation, however they do provide a number of
useful estimates of expected spectral outcomes.

Finally, the cascade takes place extremely rapidly on cosmological
scales.  Direct computation of redshift effects on cascade spectra is
not needed since from the standpoint of cascading particles the
universe is essentially flat and static.  One may however easily extend
the spectrum to produce real cosmological flux limits using
\eqref{omegalimiteqn}.

\section{Summary}

In conclustion, $\gamma$-ray energy injected by ultra high-energy
isotropic cosmological processes is observable as a spectrum of cooled
$\gamma$-rays.  These processes differ from point sources in that
point sources suffer attenuation on the cosmological medium.

The consequences of this work may be summarized as follows:

\begin{enumerate}
\item Diffuse isotropic injection processes above 10~TeV are
  constrained by experiments which observe $\gamma$-rays below 10~TeV.

\item The total input power spectrum is reprocessed leading to a
  $\gamma$-ray pile up at $E_c$.   This output spectrum gives
  total integrated limits on injection energy.
  
\item Observations of present day fluxes suffer important restrictions
  from both energy and redshift scales.  Cosmological processes that
  contribute to the EGRET diffuse flux observations must be
  significant at $z\approx 1$.  The significance of any observation of
  present day cascade radiation processes is highly suppressed by a
  factor of 1000.
\end{enumerate}

There are energy loss processes which I have not considered in these
calculations.  First the synchrotron losses may have unit order
corrections on these calculations where large magnetic fields are
present.  Next ionization energy losses in dense galactic regions may
play a key role in $\gamma$-ray attenuation.  Finally, bremsstrahlung
losses may give a measurable correction to galactic fluxes.

For all of these loss processes, the rate of loss could easily be
measured by considering the emission of well understood sources.  In
particular, through the synchrotron mechanism this may give a testable
method of determination of the intergalactic magnetic fields.

Throughout this discussion I have taken it for
granted that the photon injection was continuous and isotropic on the
cosmological scale.   Individual point sources
represent a dramatically different type of propagation problem than
what is covered in this work.   


\section{Acknowledgment}

I acknowledge my advisor John~Beacom for collaboration and helpful
suggestions.  I thank Hasan~Y\"uksel, Mandeep~Gill, Frederick~Kuehn,
Gregory~Mack and Eduardo~Rozo for interesting discussions and editorial
comments. JAC was supported by DOE Grant No. DE-FG02-91ER40690; I also
thank CCAPP and OSU for support.


\clearpage
\thispagestyle{empty}

\begin{appendix}
\section{The Spectrum of $\gamma$-rays Scattered at $E_c$}
\label{App:BerezinskySpectrum}

Berezinksy derives the spectrum of $\gamma$-rays at $E_c$ in his
textbook \cite{Berezinsky:1990}.  In the high energy limit electrons
conserve energy, $E q_e(E) = const$, above $E_c$.  The production
of inverse Compton $\gamma$-rays is the number in a logarithmic
interval $q_e(E_c)\, dE$ over the width of a logarithmic interval
$dE_e/E_\gamma$, i.e.,  $d ln E$.   The spectrum is:

\begin{equation}
\tag{$\star$}
\label{Eq:ICSFlux}
 n_\gamma = q_e(E_c) {dE_e\over E_\gamma}
\end{equation}

But as we have said, energy is conserved, therefore at $E_c$, $q_e =
q_0$.  The energy carried by an outgoing $\gamma$-ray is the fraction
liberated from the electron or positron, $E_\gamma = f E_e$, by
Eq.~\eqref{Eq:ComptonEFrac}:

$$ E_\gamma = {4\over 3} {{E_e^2 \epsilon}\over m_e^2} $$

\noindent Taking the derivative and isolating $dE_e$ we have:

$$ dE_e = {1 \over 2} {3 \over 4} {m_e^2 \over \epsilon} {dE_\gamma
  \over E_e} $$

\noindent And solving for $E_e$:

$$ E_e = m_e \sqrt{{3 \over 4} {E_\gamma \over \epsilon}} $$

\noindent Finally we can combine these relations in \eqref{Eq:ICSFlux}:

$$ n_\gamma = {1\over 2} q_0 m_e \sqrt {3\over 4\epsilon}\, {dE_\gamma
  \over E_\gamma^{3/2}} $$

\noindent As expected, at $E_c$ the spectrum of $\gamma$-rays is falling as $E^{-3/2}$.

\section{The Spectrum of $\gamma$-rays Scattered at Higher Energies}
\label{App:BerezinskyTwoStep}

Above, we present the conventional discussion of $\gamma$-ray spectra
resulting from a single cascade step.  It is trivial to extend this
discussion to include an additional step.   If the injected particle
energy is taken as $4 E_{th}$ rather than $E_{th}$ an additional
scattering step results.  Here the low energy approximation to energy
loss fraction does not apply.   In this ``middle'' energy range the
outgoing electron and positron share total energy so the loss fraction
is now:

$$f \approx {1\over 2}$$

\noindent Then the following expression trivially follows from
\eqref{Eq:ICSFlux}:

$$n_\gamma = 2 q_0 {dE_\gamma \over E_\gamma} + n_{\gamma,2}$$

\noindent In this relation $n_{\gamma,2}$ will be the flux produced by the
subsequent outgoing $\gamma$-ray produced at $E_{th}$.   Yet we have
already deduced this spectrum in App.~\ref{App:BerezinskySpectrum}.  Therefore,

$$ n_\gamma = 2 q_0 {dE_\gamma \over E_\gamma} + {1\over 2} q_0 m_e
  \sqrt {3\over 4\epsilon}\, {dE_\gamma \over E_\gamma^{3/2}} $$

\noindent The $E^{-1}$ spectrum will dominate for $E > 1$.

\end{appendix}

\end{document}